# General Principles in the Interpretation of Quantum Mechanics.

Casey Blood
Professor Emeritus of Physics, Rutgers University
Sarasota, FL
Email: CaseyBlood@gmail.com

## Abstract

The three major theoretical principles of quantum mechanics relevant to its interpretation are: (T1), linearity; (T2), invariance under certain groups; and (T3) the orthogonality and isolation of the different branches of the state vector. These three imply the particle-like properties of mass, energy, momentum, spin, charge, and locality are actually properties of the state vector; and this in turn implies there is no evidence for the existence of particles. Experimentally there is no evidence for collapse (E1) and theoretically linearity prohibits collapse. One also has the experimentally verified probability law (E2), which is found to rule out the many-worlds interpretation. The failure of these three major interpretation—particles, collapse, and many-worlds—apparently implies an acceptable interpretation must be based on perception. Rather than being a separate principle, probability follows in this interpretation from a weak assumption on perception plus the combinatorics when an experiment is run many times. This suggests a relatively simple experimental test of the "perception" interpretation.

PACS numbers: 03.65-w, 03.65.Ta

## 1. Introduction.

The theory of quantum mechanics needs an interpretation, an explanation of how its mathematics relates to our perceptions. To understand why, we need to briefly review the history of physics.

Classical physics, discovered almost entirely by Newton around 1700, was a very successful mathematical scheme for predicting the paths of the atoms that matter was presumably made of. Since it involved the flight of easily visualized particles through our familiar three-dimensional space, classical physics gave a picture of reality which was compatible with our conventional, everyday view of the world. Quantum mechanics, however, is a different story. It was discovered in the search for a mathematical scheme which would explain why the hydrogen atom radiates certain colors of light. Rather than being the brainchild of a single person, however, there were at least half a dozen major physicists who, over a period of 25 years, each discovered an essential piece of the puzzle. And with each piece, the mathematics moved farther away from our everyday view of the world.

The end result of these half dozen separate insights—the equation discovered by Schrödinger in 1926—was a scheme which is highly successful. In addition to the hydrogen atom spectrum, quantum mechanics describes a great many phenomena—all the properties of atoms and nuclei, semiconductors, lasers, the systematics of elementary particles—and it has never given a result in conflict with experiment. But its mathematics involves the mysterious wave function rather than the easily-visualized particles, so it no longer has a clear correspondence with our commonplace view of reality. Because of this, one needs an interpretation of the mathematics to understand how it relates to our perceived physical world.

## Simplifying the interpretive problem.

Unfortunately, the interpretive issue has become quite tangled over the past 80 years. There are three major interpretations and probably a dozen others, each with its own lengthy commentary, so it might seem that we would have to traverse an endless wilderness of conflicting views to understand the implications of quantum mechanics. Fortunately, there is a shortcut. Before giving an interpretation, we can first accumulate as many relevant facts as we can about what the theory itself tells us and thereby narrow down the possibilities.

There are two major (and peculiar!) properties of quantum mechanics. The first, illustrated by the Schrödinger's cat experiment in section 2, shows that the wave function of quantum mechanics gives several *simultaneously existing* versions of reality—the cat is both alive and dead at the same time. Of course we perceive only one of those possible versions of reality. But the problem is that quantum mechanics does not tell us why we perceive a single, particular version.

The second peculiar property of quantum mechanics involves probability. The theory does not tell us whether we will see a live cat or a dead cat in a given run of the Schrödinger's cat experiment; instead, through an empirically verified law separate from the rest of the mathematics, it tells us the *probability* of seeing a live cat or a dead cat. There is currently no clear understanding of this 'rolling-the-dice' aspect of the theory.

## Problems with current interpretations.

**Particles.** One way to obtain an objective reality—as opposed to the many versions of reality in quantum mechanics—is to assume the wave function is not the perceived reality; instead the perceived reality is composed of particles which are somehow 'associated with' the wave function. But it has been shown in detail in [1], with the argument sketched here in section 3, that this possibility is almost certainly incorrect; all conventional wisdom to the contrary, there is no evidence for the actual existence of particles (or strings), and there are strong arguments that they cannot exist.

**Collapse.** A second potential way out of the multiple-versions-of-reality problem is to suppose the wave function 'collapses' down to just one version; the dead cat wave function might collapse to zero, for example, leaving just a live cat (wave function) to be perceived. But it is shown in [2], with the arguments briefly reviewed here in section 4, that there is also no evidence or credible reason to believe collapse occurs.

**Everett's Many-Worlds Interpretation.** Our take-off point in section 5 is therefore to assume: (1) there are no particles and there is no collapse, so that only the wave function, with all its versions of reality, exists; and (2) the probability law holds. The first assumption essentially gives us Everett's "many-worlds" interpretation [3], or at least that part of his interpretation which does not involve probability. However, one can show that the "bare" many-worlds interpretation—no particles, no collapse, nothing exists but the wave function—is not compatible with the probability law.

## A perceiving aspect.

Thus we are led to an interpretation in which perception has an aspect outside the laws of quantum mechanics. A tentative structure for reality consistent with this conclusion is given in section 6. There is an aspect of each of us that looks in from outside the laws of quantum mechanics and perceives just one version of the wave function of the brain. In this model of existence, the probability law follows from a weak assumption about the perceptions of the observer, but it only holds for many repetitions of an experiment in which intermediate results are not observed. If each outcome is observed, one will almost certainly obtain results that, contrary to the classical view of probability, violate the usual probability law (see section 6C). This provides a relatively simple means of experimentally testing the no-particle, no-collapse view of reality.

## 2. Properties of the Wave Function/State Vector

The familiar Schrödinger equation of quantum mechanics is an equation for the wave function, so it is (on the least abstract level) the wave functions, rather than particles, which are the 'physical objects' in the mathematics of quantum mechanics. A useful visual picture of the wave function is that it is matter spread out in a mist or cloud of varying density. The Schrödinger equation determines the shape of the cloud, how it moves through space, and how it responds to other clouds corresponding to other 'particles.' The wave function of a macroscopic object like a cat or a human being, composed of billions of individual wave functions, is of course extremely complicated, but that does not prevent us from deducing its relevant general characteristics.

The actual state of affairs, however, is somewhat more abstract. Technically when discussing quantum mechanics, we should use the abstract 'state vector,' denoted by the ket notation, $|\ \rangle$, instead of the more easily visualized wave function. In the situations we shall discuss, however, the wave function description is equivalent to the use of state vectors, so it is permissible to use the more easily visualized wave function description.

## A. Schrödinger's Cat and Versions of Reality.

The world around us certainly appears to be unique, *the* physical world, upon which we all agree. But in quantum mechanics, the highly successful mathematical description of nature, there is no unique physical world; instead there are many simultaneously existing versions of physical reality. This most peculiar state of affairs is nicely illustrated by the famous Schrödinger's cat thought experiment.

As a preliminary, we consider the wave function of a single radioactive nucleus. Radioactive decay is a process in which quantum mechanics must be taken into account. And in the quantum mathematics, there is both a part of the wave function corresponding to an undecayed nucleus *and* a part corresponding to a decayed nucleus. That is, the wave function at a given time is

[the nucleus radioactively decays]
**and, simultaneously**
[the same nucleus does not decay]

Both options, both *branches* of the wave function, exist simultaneously! One cannot get around this; the successes of QM depend on it.

The Schrödinger's cat experiment is a clever and dramatic way of boosting this strange multi-reality situation from the atomic to the macroscopic level. A cat is put in a box along with a vial of cyanide. Outside the box are a radioactive source and a detector of the radiation. The detector is turned on for five seconds. If it records one or more counts of radiation, an electrical signal is sent to the box, the vial of cyanide is broken, and the cat dies. If it records no counts, nothing happens and the cat lives.

Classically, there is no problem here (unless you are a cat lover). Either a nucleus radioactively decays, the cat dies and you perceive a dead cat when you open the box; or no nucleus decays, the cat lives and you perceive a live cat. Schematically, in the classical case,

**either**
[nucleus decays] [cat dies] [you perceive a dead cat]
**or**
[no nucleus decays] [cat lives] [you perceive a live cat]

But this is not what happens in the quantum case. There, the wave function of the nucleus, the cat, and you (as the observer) is

[nucleus decays]

[cat dies]
[ version 1 of you perceives a dead cat]
**and, simultaneously**
[nucleus does not decay]
[cat lives]
[version 2 of you perceives a live cat]

There are now two full-blown, simultaneously existing versions of physical reality.  In one, there is a dead version of the cat, in the other there is a live version.  In one, version 1 of you perceives a dead cat, in the other, version 2 of you perceives a live cat so there are *two simultaneously existing versions of you*, each perceiving something different!

In reality, your perceptions will correspond to one version or the other.

Schrödinger's cat is one example of a set of general properties which follow from the agreement in all known cases between observation and the quantum mechanically predicted characteristics of the versions of reality:

**A1.** Quantum mechanics gives many *potential* versions of reality.
**A2.** In all instances where the calculations and observations can be done, there is always one and only one version whose characteristics corresponds exactly—qualitatively and quantitatively—to our physical perceptions.
**A2\*.** Quantum mechanics does not single out any version for perception.
**A3.** All observers agree on the perceived version.

Property **A2** is, in my opinion, the pivotal observation in deducing the "correct" interpretation of quantum mechanics.

## B. Linearity. Different universes.

The most important mathematical property of quantum mechanics is that it is a linear theory.  Because of this, when a state vector divides into a sum of different, orthogonal versions, each version evolves in time *entirely independently* of the other versions present.  It is as if each version is in a different universe and there can be no communication of any kind—via light, sound and so on—between the different universes. [To see this in a particular case, do a Stern-Gerlach experiment on a silver atom.  Then the total wave function is non-zero only in two non-overlapping regions of the 'location of the silver atom.'  From this one can show that the two branches of the wave function are always orthogonal and evolve entirely independently.]

**B1.** This has a bearing on why we perceive only one version of reality. The perception of different versions occurs in different universes that cannot communicate, and so we could never be *communicably* aware of the perception of more than one version.  The version of the brain perceiving one 'reality' could not communicate that perception to a version of the brain perceiving another 'reality.'

**B2.** Similarly, the division into different, non-communicating universes in quantum mechanics implies that two observers can never disagree on what they perceive.

**B3.** These two statements imply it is appropriate to apply a "superselection rule" to the various versions of reality.  Linear combinations of, say, the various states of the observer are not prohibited, but (as in the case where one has one universe with charge $2e$ and another with charge $3e$) when there is no interaction between different solutions of the Schrödinger equation, nothing—no new physical insight—is gained by considering linear combinations of the states.  Thus the "preferred basis problem," in which linear combinations of versions of the observer are considered, seems to be a red herring.

## C. The Probability Law.

In the Schrödinger's cat experiment, the quantum mechanics of the nuclear decays will tell us that the 'sizes' (technically the norms, roughly the amount of cloud material) of the two possible wave function outcomes are different.  The cat alive part of the wave function might have a size of 2/3, for example, while the cat dead part might have a size of 1/3 (so the wave function is written as $a_1$[cat alive]+$a_2$[cat dead], with $|a_1|^2$=2/3, $|a_2|^2$=1/3).  If we do the Schrödinger's cat experiment many times, there is an additional probability law in quantum mechanics which says that 2/3 of the time we will see a live cat, and 1/3 of the time we will see a dead cat.

More generally suppose the wave function contains $K$ versions of reality, designated by $i$ ($i$=1,…, $K$). Let the 'size' (norm) of version $i$ be $|a_i|^2$, with the sum of the norms adding up to 1.  Then the $|a_i|^2$ probability law says:

> **P.** If an experiment is run many times, a physical reality with characteristics corresponding to version $i$ will be perceived a fraction $|a_i|^2$ of the time.

Statement **P** is another observation that is pivotal for deducing the correct interpretation. It is well-established, but the reason why it holds is not currently known.  (See, however, section 6.)

**C1. Consistency of the Probability Law.**  Suppose we agree that the probability of perceiving event $i$ is a function of the 'size' $|a_i|^2$ so that $p_i=f(|a_i|^2)$.  Then there are any number of ways to show that the only functional form consistent with the mathematics of quantum mechanics is the conventional $f(|a_i|^2)= |a_i|^2$.  The law $f(|a_i|^2)= (|a_i|^2)^2$, for example, would give inconsistent results.  I think this observation—that the $|a_i|^2$ probability

law is the only one consistent with the rest of conventional quantum mechanics—strongly suggests that the origin of the law must, to a large extent, be *within* conventional quantum mechanics. (See section 6.)

## 3. No Evidence for Particles.

If we ignore properties **B1** and **B2**, as is often done in interpretations, one way of attempting to explain why we perceive only one of the versions of reality of quantum mechanics, and why we all agree on the version, is to suppose that the wave function is not the physical reality we perceive. Instead there is an actual, unique physical existence, made up perhaps of particles, which somehow conforms to just one of the versions, and it is that physical reality which we perceive. This, in fact, is the view of the majority of scientists and non-scientists alike. If you look in a typical modern physics text, you will find analyses of experiments—the Compton and photoelectric effects, for example—which reputedly prove particles are necessary for understanding physical existence.

But the problem with these arguments is that they do not take into account *all* the properties—well-known and not so well-known—of the wave function. If these are taken into account, then one can show (see [1] for details) that all the particle-*like* properties can be explained by properties of the wave function alone.

> Group representation theory applied to the inhomogeneous Lorentz group and the internal symmetry group implies mass, energy, momentum, spin and charge are properties of the wave function/state vector. And property **B1** implies our *perceptions of the effects* of a spread out wave function will be localized.

Thus there is no evidence for particles. That is, wave-particle duality is not a duality in the actual structure of matter; instead it is simply a duality in the *properties*—classical wave-like or classical particle-like—*of the wave function*.

We note in passing that the results of the Bell-Aspect experiment on non-local effects [2], and the Wheeler delayed-choice experiment [3] seem to imply questionable properties—instantaneous action at a distance, and the effect before the cause—if one assumes there are particles, but these properties are not necessary if only the wave function exists.

We note also that there are a number of barriers to constructing a *theory* of particles—such as that of Bohm [4]—that meshes properly with quantum mechanics [1]. The net result of no evidence and severe theoretical difficulties is that, contrary to popular belief, it is most unlikely that objective "matter" in the form of particles, or strings, or hidden variables or whatever exists. That is, physical existence seems to be constructed from the (multi-reality) wave functions alone.

# 4. No Evidence for Collapse.

Another way of attempting to explain why we perceive only one version of reality is to suppose that the wave function somehow *collapses* down to just one version. In Schrödinger's cat, for example, the dead cat version of reality might collapse to zero—that is, it would simply go out of existence—leaving just the live cat version to be perceived.

The mathematical theory of collapse proposed by Ghirardi, Rimini, Weber, and Pearle model [5-7] is analyzed in [8]. A random force is introduced into the time evolution of the wave function in such a way that, after a short period of time, for systems with many particles, there is a collapse of the wave function down to just one version (without affecting the 'shape' of the wave function); all the other versions simply go out of existence. This idea is beautifully implemented in the Ghirardi, Rimini, Weber, Pearle model. But even though the mathematics is elegant, there are difficulties with the physical implications of the scheme: there must be instantaneous coordination of billions of random events located far from each other; the linearity of quantum mechanics must be abandoned; and the specific method of collapse, using particle number, doesn't work in all situations. Further, any mathematical theory of collapse will encounter the same problems. In addition, *there is no experimental evidence for collapse* [7,8]. And so there is currently no reason to suppose that mathematically implemented collapse is the solution to the problem of perception of only one version of reality.

It has also been proposed [10-12] that collapse is induced by *conscious perception* of an event. This has the disadvantage (for me, at least) that the consciousness (a slippery concept) of a human being disrupts the mathematics. In addition, one has problems understanding how the probability law comes about. So I don't believe such collapse proposals are defendable (or at least not first choice if alternatives exist).

# 5. The Many-Worlds Interpretation Cannot Be Correct.

There is no experimental evidence for particles and there are fairly strong theoretical arguments against their existence. Further there is no experimental evidence for collapse and some theoretical evidence against its occurrence. So a reasonable goal, the one we pursue here and in section 6, is:

> *To consider the implications for the structure of reality if we assume quantum mechanics as it currently stands, with no particles and no collapse, is correct.*

To this end, we define a structure for existence, denoted by QMA, in which there is only bare-bones quantum mechanics. This is to include the linear equations of motion and the usual states of quantum mechanics. In such a system, the wave function/state vector evolves in a completely deterministic way. All versions of reality are given equal status in QMA so there is no explicitly assumed perceptual weighting factor; that is, there is no explicit probability law. We further assume that there is no 'awareness' 'outside' of quantum mechanics in QMA; that is, there is no 'Mind' or 'being' that perceives the wave function from outside physical reality. This is Everett's many-worlds scheme, without his "assumptions" on probability. We will give two arguments show that

> *QMA is not logically compatible with the $|a_i|^2$ probability law, and therefore does not give an acceptable structure for physical reality.*

### A. The probability law cannot be stated within QMA and it therefore cannot hold there.

To see this, suppose we perform a measurement on an atomic-level system with state vector $\sum_{i=1}^{n} a_i |i\rangle$. After the measurement, the state vector of the system is

$$\Psi = \sum_{i=1}^{n} a_i |O_i\rangle |A_i\rangle |i\rangle, \quad \sum_{i=1}^{n} |a_i|^2 = 1.$$

The $|A_i\rangle$ are the *n* versions of the apparatus that detect and record the *n* possible outcomes and the $|O_i\rangle$ are the *n* versions of the observer that perceive the readings on the versions of the apparatus. Each of the *n* branches constitutes a separate universe. And one can prove (see appendix A) that each version is equally conscious, equally aware.

It would seem to be straightforward to state the probability law in QMA. One possibility is:

**P1**. On a given run, the probability of the observer perceiving outcome *i* is $|a_i|^2$.

This statement is certainly correct experimentally, but it is not acceptable within the framework of QMA. "…*the* observer perceiving…" implies perception by a unique version of the observer—but there is no unique, singled-out version. Instead there are *n equally valid* versions, each perceiving a different result. So **P1** will not do. A second potential statement is:

**P2**. On a given run, the probability of perceiving outcome *i* is $|a_i|^2$.

But this dodges the issue of what it is that perceives, and that is not acceptable in this context. A third possibility, which acknowledges that it is the versions which perceive, is:

**P3**. On a given run, the probability of my perceptions corresponding to those of version $|O_i\rangle$ is $|a_i|^2$.

But because there are *n* versions of *me*, "*my* perceptions" are always associated with *every* outcome, so there is no probability. Thus **P3** will also not do in QMA (although it is quite acceptable "in reality"). Finally, we might try:

**P4**. The probability of outcome *i occurring* is $|a_i|^2$.

But that is not acceptable either, because all outcomes occur in QMA on every run. The only probability relevant to QMA is probability of *perception*.

The point is this: The probability law is about what is perceived, and the only entities that perceive in QMA are the *n* versions of the observer. So

> *The probability law in QMA must be written solely in terms of the perceptions of the versions of the observer,*

with no reference to "my" perceptions or the perceptions of "the observer" (because there is no "me" or "observer" different from the versions). But that is patently impossible because every version of the observer perceives its respective outcome on every run with 100% certainty; there is nothing probabilistic about the perceptions of the versions.

To restate the argument: When only the versions perceive, when every version is of equal perceptual status so that no version is singled out as "me," and when each version perceives its respective outcome on every run—that is, when there is no probabilistic or coefficient-dependent aspect to the *n* perception processes—there is no way within QMA to obtain a coefficient-dependent probability of perception.

One might be tempted to simply add the assumption of probability to the rules of QMA to get, say, QMB. But one cannot do even that. When only the versions perceive, when every version perceives on every run, and when each version is equally conscious then one cannot add the assumption of probability of perception because one cannot write the law for probability of perception! This is just an indirect way of saying that:

> *The rules of QMA—linearity, only the quantum versions of the observer perceive, no particles or hidden variables, no collapse—do not allow probability of perception.*

### B. Probability implies singling out.

By looking at a particular case, we can see the problem with probability in QMA from a different perspective. Suppose we consider a two-state system, $\psi = a_1|1\rangle + a_2|2\rangle$, with $|a_1|^2 = .9999, |a_2|^2 = .0001$, and suppose we do ten runs, with the observer perceiving the results of every run. There will be $2^{10} = 1,024$ possible outcomes and 1,024 versions of the observer, each equally valid. But we know experientially that my perceptions will (almost always) correspond to only one of them, the one with all ten outcomes 1. That is, one version from among all the 1,024 versions is (almost always) *singled out* as the one corresponding to my experiential perceptions.

But in QMA, every one of the 1024 versions of the observer perceives its respective outcome on every run of 10. There is no reason *within QMA*—where all versions of me are equally aware—why my perceptions should almost never correspond to any of the other 1023 versions. That is, the probability law is not consistent with the conditions imposed by QMA.

### C. The work of others.

A number of others [13-20] have claimed that probability is consistent with the many-worlds interpretation. A detailed critique is given in [21]. A blanket criticism is that none of them has given a satisfactory statement of the law of probability of perception. And several of them explicitly or implicitly assume, without proper justification, that probability is a property of the state vector.

Several of these treatments use an 'auxiliary experiment' methodology which, it is claimed, shows that the probability law follows from certain assumptions. We argue in ref. [21], sec. 4, however, that at least one of the assumptions cannot be justified.

## 6. The Mind Interpretation.

We have shown that QMA does not give an adequate framework for accounting for our perceptions (including the probability law). It therefore seems reasonable to go further and explore how QMA needs to be modified to yield a scheme that gives a proper account. Two possibilities are to assume that QMA is incorrect in the sense that either particles exist or there is collapse. But because there is no evidence for either of these, we will here consider no-particle, no-collapse structures for reality (see also [21]).

Assuming no-particles and no-collapse, we know from the failure of QMA that awareness cannot be based *solely* in the physical—that is, quantum mechanical—brain (although the brain is certainly heavily involved in our perception of the world). Thus something must be added to QMA and that 'something' must presumably pertain to awareness. To obtain some idea of what is needed, we repeat property **A2** here:

> **A2.** In all instances where the calculations and observations can be done, there is always one and only one version whose characteristics corresponds exactly—qualitatively and quantitatively—to our physical perceptions.

This tells us that existence is constructed *exactly as if* there were a 'Mind' looking in from outside physical reality—that is, from outside the rules of quantum mechanics—and perceiving just one quantum version of the wave function of our brain. And so this is what we will assume in the Mind interpretation of quantum mechanics, abbreviated QMM. Its properties are given in items 1-6 below. This assumption of a unique 'I,' as opposed to the many versions of 'I' in QMA, has the advantage that it matches our intuitive view of ourselves (but of course the intuitive view cannot be used as scientific evidence for the Mind.)

## A. Basics of the Mind Interpretation of Quantum Mechanics

**1. The non-physical Mind.** Associated with each individual person is an individual Mind that is not subject to the mathematical laws of quantum mechanics. In particular, the Mind has no wave function/state vector associated with it. We use a capital M to distinguish this Mind from the usual usage of the word mind. Because quantum mechanics describes the physical world so well, we will *define* anything outside its laws to be 'non-physical.' So with this definition, the Mind is non-physical.

**2. The Mind perceives only the brain-body.** The individual Mind perceives only the wave function of the individual brain (brain-body); it does not directly perceive the quantum state of the external world. This is an acknowledgement of the fact that we are not directly aware of the external world; we are only aware of the neural state of our brain. The process of perception of the wave function/state vector by the Mind is not understood.

**3. The Mind picks out only one version of the wave function.** The individual Mind concentrates on one version of the wave function of the brain-body, and it is the concentrated-upon version that enters our conventional awareness.

**4. No collapse.** The Mind does not collapse the wave function or interfere with the mathematics of QMA in any way.
> A primary objection to dualism—a non-physical Mind separate from a physical brain-body—is that the non-physical aspect must exert a force or otherwise has some effect on the physical. The QMM scheme circumvents this objection because the non-physical aspect only *perceives*; it does not affect the physical world (which is made up of wave functions/state vectors) in any way.

## B. Agreement among Observers. The Overarching MIND.

The model as it has been given so far leaves two important questions unanswered—why observers agree on what they perceive, and why the probability law holds. To make the first question specific, consider again the Schrödinger's cat experiment and suppose we have two observers. Then according to the rules of quantum mechanics, the wave function is

[cat alive]
[obs 1's brain state corresponds to cat alive]
[obs 2's brain state corresponds to cat alive]

*—and—*
[cat dead]
[obs 1's brain state corresponds to cat dead]
[obs 2's brain state corresponds to cat dead]

Suppose observer 1's Mind focuses on the version of the associated brain corresponding to cat alive so that observer 1 perceives, in the everyday sense, a live cat. We know from everyday experience that observer 1 and observer 2 (and the cat) must be in agreement. And we know from property **B2** in section 2 that two observers can never disagree. But still, how do we guarantee in our Mind model, that observer 2's Mind is also focused on the "live" version of the brain? There is a way to bring about agreement but it is bound to make scientists even more skeptical of this proposal because it is far outside the realm of traditional science.

> **5. The overarching MIND.** Instead of each individual Mind being separate from all others, each Mind is a fragment or facet of a single overarching MIND. Each individual Mind is that aspect of MIND that is responsible for perceiving the state of the associated individual physical brain. Perception of a particular version of the wave function by one individual Mind is then presumed to set the perception of that same version by the overarching MIND. And that in turn sets the perception of the same version by all the other individual Minds.

So to obtain agreement among observers in this scheme, we are apparently forced to substitute the MIND assumption for the conventional particle or collapse assumptions that give a single-version physical world. The scientist will say there is no evidence to justify such an outrageously non-scientific assumption. But the counter-argument is that there is no evidence to justify the particle or collapse schemes either. And if there is no collapse and there are no particles—the option we are exploring here—we are certainly forced outside QMA. It then seems to me that **A2** pretty much forces us to the Mind-MIND scheme of QMM.

## C. The Probability Law.

The astute reader may have realized that we have a problem with the concept of probability if we assume no particles and no collapse. Probability as it is traditionally understood, say in the dice-rolling example, refers to the probability of a specific, actual, 'single-version' event; there is an actual die that has a specific reading after each roll. But in quantum mechanics, the wave function does not give a specific, actual, single-version event; instead, it gives several versions of reality (equivalent to all six readings of the die occurring at the same time), each potentially corresponding to an 'actual, perceived' event. So the direct use of classical probability is not appropriate in QMM.

In spite of this, however, there is a way to salvage the probability law in the QMM scheme. To do this, we make the following (relatively weak) assumption:

> **6. Probability.** The individual Mind (or the overarching MIND) is 'much more likely' to perceive a version of reality that has a much larger norm than other versions. The Mind doesn't have to follow the usual $|a_i|^2$ probability law, or any stable probability law at all; it just has to be much more likely to perceive those versions with a large norm.

To see how this leads to the $|a_i|^2$ probability law, suppose we do a spin ½ Stern-Gerlach experiment $N$ times, with $N$ large. The amplitude for spin (+ ½) is $a_1$ and that for spin (– ½) is $a_2$. Then if the observer observes only the end result—and not the intermediate ones—the amplitude squared for perceiving $m$ (+ ½) spins and $N - m$ (– ½) spins is

$$|A(m,N)|^2 = \frac{N!}{m!(N-m)!}|a_1|^{2m}|a_2|^{2N-2m}$$

A simple calculation using Stirling's approximation for the factorials shows that the amplitude squared for $m$ near the max at $m=N|a_1|^2$ is much larger than the amplitude squared when $m$ is not near $N|a_1|^2$. Thus according to property 6, the *perceived* result will have $m$ near $N|a_1|^2$. (See also appendix B.)

> It is worth emphasizing two points here.
> • The probability law only pertains to a large number of repetitions of the experiment (which is, of course, pretty much true in classical probability theory also).
> • The probability law for long runs only holds if individual results are *not* observed (which is quite different from classical probability theory).

The second point is most interesting because it implies this view of the probability law can be experimentally tested. Suppose the Stern-Gerlach (or some other) experiment is carried out in such a way that the observer perceives the outcome of *every* run. Then if the above holds, it is possible, indeed likely, that after many runs of the experiment, the average value of $m/N$ could be relatively far from $|a_1|^2$. For further comments on this possibility, see appendix B.

# 7. Summary.

**Failure of the particle, collapse, and many-worlds interpretations.**
Quantum mechanics contains a most interesting paradox. It always (where we can check) gives the correct answer, but at the same time, it contains several simultaneously existing versions of reality instead of the single version we perceive.  One potential way out of this dilemma is to suppose the perceived reality consists of particles rather than wave functions.  But it can be shown that there is no evidence for particles, and very little chance they exist.  A second potential way out is to assume the wave function collapses down to just one version.  But we have also noted that there is no experimental evidence for collapse, and there are substantial theoretical obstacles—including linearity—to the construction of a theory of collapse.  Thus it seems reasonable to explore the consequences of assuming there are no particles and no collapse.

If we make these assumptions, then physical reality consists solely of the wave function with all its versions of reality.  That is, we essentially have the many-worlds interpretation.  But one can *prove* that many-worlds is not a possibility, because the experimentally verified $|a_i|^2$ probability law cannot hold.  Thus, under quite reasonable assumptions, we see that the three major interpretations of quantum mechanics don't work.

### The Mind-MIND interpretation.

How then are we to interpret quantum mechanics?  Our proposal is that, associated with each of us is a Mind that perceives just one version of the wave function of our individual brain, with this Mind not being subject to the mathematical laws of quantum mechanics.  To make certain every observer is 'consciously' aware of the same version, we must further assume that there is an over-arching MIND, and that our individual Minds are aspects of that MIND.  This Mind-MIND interpretation is certainly outside the realm of current science and runs contrary to a 'materialist' view of existence.  But since (under our assumptions) the three major interpretations don't work, it seems to be the most likely structure for existence.  (See appendix C for comments on the Copenhagen interpretation in this regard.)

The Mind-MIND interpretation has the advantage that the highly successful mathematics of quantum mechanics need not be altered at all.  In addition, in contrast to proposals advocating dualism in *classical* physics, the Mind in this quantum-based duality scheme does not tamper with or alter physical reality (the wave function) in any way; it merely *perceives*.

### Probability.

If there are no particles and there is no collapse, that is, if there are many versions of reality instead of just one, then the idea of probability must be reconsidered.  It is noted in section 6 that if we have a perceiving aspect—a Mind—that satisfies a certain weak condition, the $|a_i|^2$ probability law holds for a large number of repetitions when intermediate results are not perceived.  But if individual results are perceived, it is quite possible the $|a_i|^2$ probability law will not be followed.  So a test of the probability law in which all outcomes are perceived should provide an indirect test of the no-particle, no-collapse scenario.  If all outcomes are perceived and the probability still follows the $|a_i|^2$ law, then the no-particle, no-collapse, non-physical Mind scheme must be rethought. But if, contrary to all current expectations, the $|a_i|^2$ law is not followed when all outcomes are perceived, then the non-physical Mind scheme is almost certainly correct.

## Appendix A. All Versions Are Equally Aware.

To show that all versions of the observer are equally aware, we perform a measurement on an atomic-level system with state vector $\sum_{i=1}^{n} a_i | i \rangle$. After the measurement, but before the observer looks at the reading on the detector, the state vector of the system is

$$\Psi = | O_0 \rangle \sum_{i=1}^{n} a_i | A_i \rangle | i \rangle, \ \sum_{i=1}^{n} |a_i|^2 = 1.$$

The $| A_i \rangle$ are the $n$ versions of the apparatus that detect and record the $n$ possible outcomes. After the observer perceives the readings on the detector, the state vector is

$$\Psi = \sum_{i=1}^{n} a_i | O_i \rangle | A_i \rangle | i \rangle, \ \sum_{i=1}^{n} |a_i|^2 = 1.$$

Our perceptions correspond exactly to those of one of the versions of the observer, so one might naively expect that just one version evolves to "conscious awareness." One can show, however, that that is incorrect in QMA; all versions evolve to the aware state.

To see this, we first note that the only factors which could influence "awareness" in the no-collapse, no-hidden-variable, no-outside-mind MWI are the *characteristics of the state vector* of the brain (brain-body) of the observer. Second, the $n$ evolutions of the state vector $| O_0(t_a) \rangle \to | O_i(t_b) \rangle$ from time $t_a$, after the measurement is made but before the observer looks at the reading on the apparatus, to time $t_b$, after the observer looks, take place *separately* on each branch (because of linearity and orthogonality). If the evolution from an aware state to either an aware or an unaware state were probabilistic—perhaps from the differing influences of the environment or the internal states of the brain—then in some instances, the $n$ separate probabilistic evolutions will yield *no* aware versions, and that would give an unacceptable interpretation. The only way to avoid the no-aware-version problem is to assume that *all* versions evolve from aware to aware. Thus if "*I*" perceive one result, there are $n - 1$ other, equally aware "*I*"s perceiving the other results.

## Appendix B. Comments on Probability.

### Non-Standard Probability.

How could the perceived results of the Stern-Gerlach experiment always be near $m/N=|a_1|^2$ if just the end result is perceived, but not near that value if every intermediate result is perceived? It does *not* arise because the non-physical Mind is influencing the outcome of each event (there is, in fact, no 'outcome of each event' because all the versions of reality continue forever). To see what is happening, suppose the 'probability law' for the Mind perceiving event $i$ is $s(x) = x + .1(x)(1-x)(.5-x)$ where $x$ is the amplitude squared. Then if each individual result is observed, the value for $m/N$ will be near $s(x), x = |a_1|^2$, in disagreement with the $|a_i|^2$ probability law. But if only the end result is observed, the value for $m/N$ will be the one that maximizes $s(x), x = |A(m,N)|^2$ where the $A$ is given in section 6C. From the chain rule for derivatives, this will give a maximum at $m/N = |a_1|^2$, in agreement with the $|a_i|^2$ probability law.

### The Rutherford Experiment.

One might suspect that experiments in which every outcome is observed have already been done. In particular, suppose we consider the original Rutherford scattering experiment, where a flash of light was observed (by a graduate student) every time a decay particle hit the zinc sulfide detector. This experiment won't do as a test of the probability law, however, because not *every outcome* is observed. Only that very small fraction of events where the decay particle hits the small detector is observed. And one can show in that case that the $|a_i|^2$ law will indeed hold (except perhaps for an overall normalization factor) no matter what the 'probability law' is for the perception of individual events.

To sketch the argument, suppose we do a scattering experiment in which there are R scattering events per second, but we observe only those events which register in a very small solid angle. Then one can prove that the scattering amplitudes squared are, to a good approximation, functions of the product $Rt|a_i|^2$ rather than being functions of $|a_i|^2$ and $Rt$ separately, where $t$ is the time the experiment has been run, and $a_i$ is the amplitude for scattering into solid angle $i$. This implies that, *independent of the probability law for perception*, the average time for observing a single event $i$ is proportional to $1/(R|a_i|^2)$, which is exactly the result one obtains if the probability law is $|a_i|^2$. Thus the individually observed scattering events in the Rutherford experiment give no information on the specific probability law.

## Appendix C. The Copenhagen Interpretation.

In opposition to our QMM interpretation, the scientific materialists might invoke a variation of the Copenhagen interpretation in which it is maintained that, because we don't directly perceive anything other than our macroscopic world, we have no business inferring anything about the true nature of reality—including non-physical awareness—from quantum mechanics. I just don't buy this no-inference-allowed argument. We can mathematically establish certain properties of the wave function. And if what we perceive always matches those mathematical properties—we observe the mathematically predicted frequencies of light from the hydrogen atom, for example—then it seems legitimate to infer that physical reality is made up of wave functions. *The ubiquitous agreement between our observations and the mathematics of the atomic world is a very powerful argument for the 'existence' of the atomic world.*

In fact, I think the Copenhagen argument works better in reverse. What we are *directly* aware of corresponds solely to our *neural representation* of reality. But experimental and mathematical physics gives a much more detailed, in-depth, and unified picture of the physical world than our sense-based view. Thus in my opinion, the picture of the physical world that mathematical physics gives us is more likely to correspond to a scheme which is closer to the 'actual nature' of reality than the one given by our neural representation.

One more point: The materialist and others might argue that there are so many seemingly valid interpretations that it is not possible to glean an accurate, consistent picture of reality from quantum mechanics. But that is not true. The choice is clear: either you find evidence for particles; or you find evidence for collapse; or you must accept that awareness is not based in the physical brain. That is, 'awareness is not based in the brain' is not an interpretation; if there are no particles (hidden variables) and no collapse, it is a *requirement*.

# References.

[1] Casey Blood, *No Evidence for Particles*, arXiv, quant-ph/ 0807.3930 (2008).
[2] A. Aspect, P. Grangier, and G. Rogers, *Phys. Rev. Lett.* **47**, 460 (1981).
[3] V. Jacques *et al*, *Science* **315**, 966 (2007).
[4] David Bohm, *Phys. Rev*. **85,** 166, 180 (1952). D. Bohm and B. J. Hiley *The Undivided Universe* (Routledge, New York, 1993).
[5] G. C. Ghirardi, A. Rimini and T. Weber, Phys. Rev. D**34**, 470 (1986); Phys. Rev. D**36**, 3287 (1987).
[6] Philip Pearle, arXiv, quant-ph/0611211v1 (2006).
[7] Philip Pearle, arXiv, quant-ph/0611212v3 (2007).
[8] Casey Blood, *Difficulties with Collapse Interpretations of Quantum Mechanics*, arXiv, quant-ph/ 0808.3699 (2008).
[9] Hugh Everett, III, *Reviews of Modern Physics,* **29**, *454* (1957).
[10] Eugene P. Wigner, in I. J. Good (ed.), *The Scientist Speculates: An Anthology of Partly-Baked Ideas*, London: Heinemann (1962).
[11] Stapp, Henry P. (1993). *Mind, Matter, and Quantum Mechanics*, Springer Verlag, Berlin.
[12] Stapp, Henry P. (2007). *Mindful Universe*, Springer Verlag, Berlin.
[13] Lev Vaidman, *On schizophrenic experiences of the neutron or why we should believe in the many-worlds interpretation of quantum theory.* International Studies in the Philosophy of Science, Vol. 12, No. 3, 245-261 (1998).
[14] Lev Vaidman, *Many-Worlds Interpretation of Quantum Mechanics.* Stanford Encyclopedia of Philosophy. (http://plato.stanford.edu/entries/qm-manyworlds/)

(2002).
[15] David Deutsch, "Quantum Theory of Probability and Decisions", *Proceedings of the Royal Society of London* **A455** 3129-3137 (1999).
[16] David Deutsch, *Quantum theory of probability and decisions,* arXiv:quantph/9906015 (1999).
[17] David Wallace, *Quantum probability from subjective likelihood: improving on Deutsch's proof of the probability rule*, arXiv:quant-ph/0312157v2 (2005).
[18] Simon Saunders, *Derivation of the Born rule from operational assumptions*, arXiv:quant-ph/0211138v2 (2002).
[19] Simon Saunders and David Wallace, *Branching and uncertainty*, British Journal for the Philosophy of Science, **59**(3):293-305 (2008).
[20] David Wallace, *The quantum measurement problem: state of play*, arXiv:quant-ph/0712.0149v1 (2007).
[21] Casey Blood, *Problems with Probability in Everett's Interpretation of Quantum Mechanics,* arXiv:quant-ph/0901.0952v2 (2009).